\documentclass{aa}
\usepackage{graphics,epsfig,txfonts}
\usepackage{enumerate}
\usepackage{rotating}
\usepackage{lscape}
\usepackage{longtable} 
\usepackage{natbib}
\bibpunct{(}{)}{;}{a}{}{,}

\newcommand{\ergcm}[1]{erg cm$^{-2}$ s$^{-1}$}

\newcommand{\ergs}[1]{$\times 10^{#1}$ erg s$^{-1}$}

\newcommand{\ltsima}{$\buildrel < \over \sim$}
\newcommand{\lsim}{\lower.5ex\hbox{\ltsima}}
\newcommand{\gtsima}{$\buildrel > \over \sim$}
\newcommand{\gsim}{\lower.5ex\hbox{\gtsima}}
\newcommand{\xmm}{{\it XMM-Newton}}

\newcommand{\asca}{{\it ASCA }}

\newcommand{\src}{\hbox{AX\,J0043-737}}

\usepackage{color}

\begin{document}
 
\title{Probing the nature of AX J0043$-$737: Not an 87 ms pulsar in the SMC}

\author{     C. Maitra\inst{1,2} 
    \and     J. Ballet\inst{2} 
    \and     P. Esposito\inst{3}  
     \and    F. Haberl\inst{1}
     \and    A. Tiengo\inst{4,5,6}
     \and    M.~D.~Filipovi\'c\inst{7} 
     \and    F. Acero\inst{2}
       }

\titlerunning{AX J0043$-$737}
\authorrunning{Maitra et al.}

\institute{Max-Planck-Institut f\"{u}r extraterrestrische Physik, Giessenbachstra{\ss}e, 85748 Garching, Germany
        \and Laboratoire AIM, CEA/DRF - CNRS - Universit\'{e} Paris Diderot, IRFU/DAp, CEA-Saclay, 91191 Gif-sur-Yvette, France
	\and Anton Pannekoek Institute for Astronomy, University of Amsterdam, Postbus 94249, NL-1090 GE Amsterdam, The Netherlands
	\and
         Scuola Universitaria Superiore IUSS Pavia, piazza della Vittoria 15, I-27100 Pavia, Italy
         \and
        INAF, Istituto Nazionale di Astrofisica, IASF-Milano, via E. Bassini 15, I-20133 Milano, Italy
        \and
        INFN, Istituto Nazionale di Fisica Nucleare, Sezione di Pavia, via A. Bassi 6, I-27100 Pavia, Italy
        \and
        Western Sydney University, Locked Bag 1797, Penrith South DC, NSW 1797, Australia
	   }


 \abstract
{} 
{AX J0043$-$737 is a source in the \asca catalogue, the nature of which is uncertain. It is most commonly classified as a Crab-like
pulsar in the Small Magellanic Cloud (SMC) following apparent detection of pulsations at $\sim$ 87 ms from a single \asca observation. A follow-up \asca observation
was not able to confirm this, and the X-ray detection of the source has not been reported since.}
{With a dedicated \xmm\, observation, we studied the nature of the source. We ascertained the source position, searched for the
most probable counterpart and studied the X-ray spectrum. We also analysed other archival observations with the source in the
field of view to study its long-term variability.}
{With the good position localisation capability of \xmm\,, we identify the counterpart of the source as 
MQS J004241.66--734041.3, an AGN behind the SMC at a redshift of 0.95. The X-ray spectrum can be fitted with an absorbed power law
with a photon-index of $\Gamma=1.7$ which is consistent with that expected from AGNs. By comparing the current \xmm\, observation with an archival \xmm\ and two
other \asca observations of the source, we find signatures of long-term variability which is also a common phenomenon in AGNs. All of the above
are consistent with AX J0043$-$737 being an AGN behind the SMC.}
{}

\keywords{galaxies: Magellanic Clouds --
          quasars: general --
          X-rays: galaxies --
          pulsars: AX J0043$-$737 --}

\maketitle

 \section{Introduction}
\label{sec:introduction}
AX J0043$-$737 is an elusive source in the \asca catalogue that deserves attention. The nature of the
source is still uncertain, even 17 years after its first detection. The source, alternately known
as SXP0.09 is most commonly classified as a Crab-like pulsar in the Small Magellanic Cloud (SMC), although an X-ray binary or an Active Galactic Nucleus (AGN)
behind the SMC have also been suggested
 as the possible nature of the source. 
 
AX J0043$-$737 was discovered during \asca observations of the SMC on 10-11 May 1999 \citep{2000IAUC.7361....2Y,2003PASJ...55..161Y}. Pulsations at  87.58073 $\pm$  0.00004 ms 
were reported from the source making it the fastest pulsar in the SMC. The significance of the detection was however marginal at 99.98\%. A followup
\asca observation of the source with a longer exposure could not confirm the pulsations. A possible explanation was the lower count rate and a 
larger background fraction in the latter observation. Confirmation of the pulsations with a higher significance and hence inferring the nature of the source was thus highly
solicited. The X-ray spectrum of the source was consistent with an absorbed power-law of photon index ($\Gamma$) of $\sim$ 1.7. The inferred X-ray luminosity at the distance of
SMC (60 kpc) was 8.6 \ergs{34}.

The X-ray detection of the source has not been reported ever-since. \src\, is still quoted in the literature of pulsars in the SMC as a possible Crab-like pulsar, 
the fastest in the SMC \citep[e.g.:][]{2004ApJ...609..133M,2005MNRAS.356..502C,2005A&A...442.1135L,2008ApJS..177..189G,2011MNRAS.413.1600R,2016ApJ...829...30C}.
In addition no radio-continuum source was ever found in the {\it ATCA} surveys \citep{2001ApJ...553..367C,2012SerAJ.184...93W}.
In this paper we report a dedicated \xmm \,observation of \src. With the precise position localisation capability of \xmm\, and its good sensitivity, we classify \src\, as an AGN behind the SMC. This breaks the decade long
mystery on the nature of the source. Using other archival observations we also study its long-term variability.
Sect. 2 describes the observations and data reduction, Sect. 3 the results and discussion and Sect. 4 the conclusions and summary. 
\begin{table*}[ht]
\caption{Observations of AX J0043$-$737}
\begin{tabular}{c c c c c c}
\hline \hline
Observation & ObsID & Exposure & Date & Datamode & Off-axis \\
 &  & ks & dd/mm/yyyy &  & \\
\hline
ASCA1 & 67021000 & 20 & 10/05/1999 & Pulse Height (GIS) & 2.7$' $\\
ASCA2 & 48003000 & 60 & 04/07/2000 & Pulse Height (GIS) & 4.3$'$\\
XMM1 & 0301170301 & 15 & 04/06/2006 & FULL FRAME (PN) & 5$'$  \\
XMM2 & 0764780201 & 40 & 16/10/2015 & SMALL WINDOW (PN) & 0$' $\\

 \hline
\end{tabular}\\
\label{table1}
\end{table*}

\section{Observations and data reduction}
\label{sec:observations}
AX J0043$-$737 was observed with \xmm\, on 16 October, 2015 (OBSID:0764780201) for 40 ks. In addition we found an  archival {\it XMM} 
observation where the source was in the field of view, and also analysed the two \asca observations of the source. The
details of the observations are given in Table~\ref{table1}.

\xmm/EPIC \citep{2001A&A...365L..18S,2001A&A...365L..27T} observations were processed with the latest \xmm\, data analysis software SAS 
version 16.1.0\footnote{Science Analysis Software (SAS): http://xmm.esac.esa.int/sas/}.
Examination for periods of high background flaring activity was performed by extracting light curves in the energy range of 7.0-15.0 keV
and removing time intervals with background rates $\geq$ 8 and 2.5 cts ks$^{-1}$ arcmin$^{-2}$ for EPIC-PN and EPIC-MOS respectively \citep{2013A&A...558A...3S}.
 Event extraction was performed using the SAS task \texttt{evselect} applying the standard filtering criteria (\texttt{\#XMMEA\_EP \&\& PATTERN<=4} for EPIC-pn and \texttt{\#XMMEA\_EM \&\& PATTERN<=12} for EPIC-MOS).

\asca observations \citep{1994PASJ...46L..37T} were analysed following standard procedures\footnote{https://heasarc.gsfc.nasa.gov/docs/asca/abc/abc.html/}.
Only data from the gas-scintillation imaging proportional counters (GIS2 and GIS3) were used for this work, as the source fell in a CCD gap for the X-ray CCD detectors (SIS).
Spectra were extracted from the screened events using the FTOOLS task {\it xselect} after applying standard filtering criteria. 


\section{Results and discussion}
\label{sec:analyses}
\subsection{X-ray position}
A maximum-likelihood source detection analysis was performed on the \xmm/EPIC images to determine the X-ray position of AX J0043$-$737.
Fifteen images were created from the EPIC cameras in five energy bands as given: $1\rightarrow(0.2-0.5)$ keV, $2\rightarrow(0.5-1.0)$ keV, $3\rightarrow(1.0-2.0)$ keV, $4\rightarrow(2.0-4.5)$ keV, $5\rightarrow(4.5-12.0)$ keV \citep{2009A&A...493..339W,2013A&A...558A...3S}.
Source detection was performed simultaneously on all the images using the SAS task {\tt edetect\_chain}. 
Astrometric boresight correction was performed with the task {\tt eposcorr} accounting for a linear shift. The reference sources 
were a catalogue of background AGNs with known redshifts as well as sources selected 
using ALLWISE mid-infrared colour selection criteria \citep{2012MNRAS.426.3271M,2015ApJS..221...12S}. 

From the two \xmm\, observations the error-weighted mean of the position was 
R.A. = 00$^{\rm h}$42$^{\rm m}$41\fs67 and Dec. = $-$73\degr40\arcmin40\farcs8 (J2000)
with a $1\sigma$ statistical uncertainty of 0.27\arcsec. 
The positional error is usually dominated by systematic astrometric uncertainties. Therefore, we added
a systematic error of 0.37\arcsec\ in quadrature \citep{2016A&A...590A...1R}. The improved position of the source is $\sim$ 0.3 arcmin
away from the original value as found from the \asca observations. With the new position we found that AX J0043$-$737 is coincident with XMMU J004241.5$-$734041
in the \xmm\, point source catalogue of the SMC \citep{2013A&A...558A...3S}, and XMMU J004241.5$-$734039 in the 3XMM-DR6 catalogue \citep{2016yCat.9050....0R}. 

\subsection{Source identification and search for counterpart}
\begin{figure}
\centering
\includegraphics[scale=0.33]{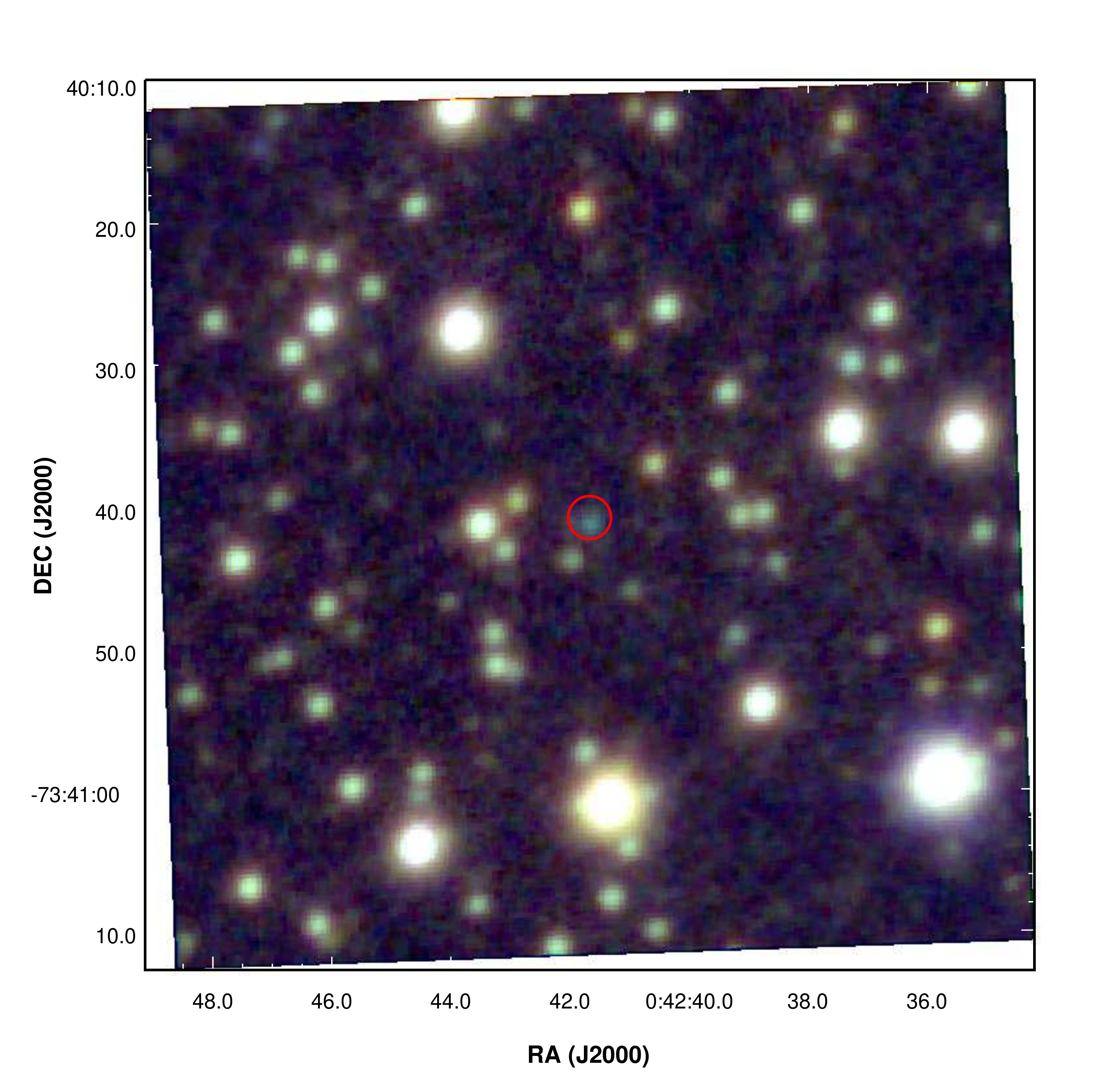}
\caption{VISTA image of the region around AX J0043--737. The image covers 1\arcmin\ x 1\arcmin\ sky area. The  
RGB colours correspond to the J, H, and K VISTA filters respectively. The `red' circle is centred on the \xmm~ position
of AX J0043--737 with a radius corresponding to 3.439$\sigma$ confidence error circle (including statistical and systematic errors.)}
\label{vista}
\end{figure}
XMMU J004241.5$-$734041 is classified as an AGN candidate in the SMC point source catalogue from its optical and X-ray colours \citep{2013A&A...558A...3S}. 
Recently the all-sky catalogue of 1.4 million AGNs of \cite{2015ApJS..221...12S}, and the Half-Million Quasars catalogue (HMQ)
and MILLIQUAS catalogue of \cite{2015PASA...32...10F,2017yCat.7277....0F} have increased the number of AGNs behind the SMC by several orders of magnitude.
To verify the credibility of the classification of AX J0043$-$737 as an AGN candidate, we correlated the source with the above mentioned catalogues with an angular separation of 
      \begin{equation} 
       r \leq 3.439 \times  \sqrt{\sigma_{\rm X}^2+\sigma_{\rm catalogue}^2} = 3.439\sigma.
      \end{equation}
For a Rayleigh distribution this corresponds to a 3$\sigma$ completeness. We found the most probable counterpart (best match) as MQS J004241.66-734041.3 \citep[catalogued in][]{2015PASA...32...10F,2017yCat.7277....0F} 
within 0.49\arcsec of the source position. To calculate the probability of chance coincidence we calculated the total number of AGNs identified in X-rays from the AGN catalogues within the \xmm\, survey area of the SMC.
This resulted into a total of 270 sources identified within 6.67 deg$^2$. 
Considering the area of the error circle corresponding to equation 1, and that the AGNs are distributed homogeneously within the survey area,
the probability of chance coincidence is $8.4\times10^{-6}$.

MQS J004241.66$-$734041 was spectroscopically confirmed by 
\cite{2013ApJ...775...92K}. \cite{2013ApJ...775...92K} selected candidates from the Optical Gravitational Lensing Experiment (OGLE-III) based on their
optical variability and
/or X-ray properties (OGLE-III ID of the source: smc128.4.5190). The $V$, $R$ and $I$ magnitudes of the source are 20.18, 19.99 and 19.91 respectively.
The redshift is $z=0.95$ indicating a luminosity distance of 6.2 Gpc to the AGN assuming standard cosmological parameters. 

In addition, we also looked for a possible counterpart in the Magellanic Clouds Photometric Survey (MCPS) \citep{2002AJ....123..855Z}.
The closest optical counterpart has $V$ and $I$ magnitudes consistent with MQS J004241.66$-$734041.3. There is no other bright and statistically
acceptable optical counterpart. This is further ascertained in Fig.~\ref{vista}, which shows the finding chart for AX J0043$-$737. This was constructed 
using the J, H and K band VISTA images which are publicly available\footnote{http://horus.roe.ac.uk/vsa/index.html}.
Considering these we conclude that
AX J0043$-$737 is the X-ray counterpart of MQS J004241.66-734041.3, an AGN behind the SMC.
\subsection{Timing analysis}
\label{sec:time}
The PN data of XMM2 were taken in small window mode, providing a time
resolution of 5.7 ms.
We searched the PN data in the 0.3--10 keV energy range for possible
periodic signals using a Fourier transform in the frequency range spanning $\sim 0.00002-88.2$ Hz. No significant signal was
found, with a 3$\sigma$ upper limit on the pulsed fraction of
$\sim$40\%\ (assuming a sinusoidal profile). We also performed a
search with the $Z^2$ test in a small interval bracketing the
\emph{ASCA} tentative detection (85--90 ms). The most prominent peak
occurs at $87.53022(2)$\,ms, with a $8\times10^{-7}$ single-trial
probability. When the $\sim$$6\times10^4$ independent period searches
are taken into account, its statistical significance is rather low
($\lesssim$2$\sigma$). Since the significance of this signal entirely
relies on the marginal ASCA1 detection, which was not confirmed
by the ASCA2 data set, we conclude that there is no evidence in
the PN data for any periodic signal, not even restricting the search
around the \emph{ASCA} value.
\subsection{Spectral analysis}
\label{sec:spec}
For the \xmm\, observations circular regions with radii of 20\arcsec\, and annular regions with inner and outer radii of 40\arcsec\, and 60\arcsec\, were used for the source and background extraction respectively. 
The SAS tasks \texttt{rmfgen} and \texttt{arfgen} were used to create the redistribution matrices and ancillary files for the
spectral analysis. The spectra were binned to achieve 25 counts in each bin and not to oversample the intrinsic energy resolution by a factor larger than 3.

For the spectral extraction of the 2 \asca observations, a different prescription was required owing to the 
 poor angular resolution of the \asca X-ray telescopes (XRT). The PSF of XRT has a 
 relatively sharp core (FWHM of $\sim 50$\arcsec) but broad wings (with half-power diameter of 3\arcmin).
 The PSF is dependent on the energy, the angle from the optical axis as well as the azimuthal angle of the source position.
 The GIS in addition has its own PSF, which is comparable in width to that of the XRT albeit with a different shape leading to
 an additional spatial broadening\footnote{https://heasarc.gsfc.nasa.gov/docs/asca/abc/node11.html}.

 Following this, a circular region of radius 4\arcmin\, was used to extract the source spectrum.
 A large annular region with inner and outer radii of 8\arcmin\, and 10\arcmin\, surrounding the source was used to extract the background spectrum.
 However the images extracted from the combined GIS (GIS2 and GIS3)
 showed the presence of another source within the extraction region of AX J0043$-$737. The contaminating source
 was identified as XMMU J004207.7$-$734503, a candidate HMXB in the SMC point source catalogue \citep{2013A&A...558A...3S}. 
 This prompted us to fit the spectra of the two sources simultaneously in the \asca observations, after taking into account the
 contamination from one source into the other.
 In order to do this, we extracted the spectra from both sources using smaller radii to minimise the contamination (2\arcmin).
 Furthermore we accounted 
 for the percentage spillover from
 each source to the other within the source extraction regions in the following manner:
 We modeled the \asca PSF at the source positions
  from \asca observations of Cygnus X-1 taken at different positions w.r.t. the optical axis. These are available in
 the energy range of 1--10 keV at intervals of 1 keV\footnote{https://heasarc.gsfc.nasa.gov/docs/asca/abc/node11.html}.
 In order to take into account the energy dependence, we reconstructed the PSF at the position of each source by adding 
 the PSFs in different energy bands weighed by the spectrum of the respective sources. For AX J0043$-$737 a power-law with $\Gamma=1.7$ was used. For the HMXB candidate XMMU J004207.7-734503, $\Gamma=1.06$ was used. This was obtained by fitting the spectrum extracted from XMM1 with an
 absorbed power-law model (Table~\ref{table2}). Comparing the PSFs resulted in a total spillover fraction of 11.06\% from 
 XMMU J004207.7$-$734503 into AX J0043$-$737 and 9.76\% from AX J0043$-$737 into XMMU J004207.7-734503 in the energy range of 0.8-7 keV within
 the respective source extraction regions. The redistribution matrices for the GIS data were obtained from the HEASARC calibration database\footnote{ftp://legacy.gsfc.nasa.gov/caldb/data/asca/gis/cpf/95mar06/gis2v4\_0.rmf,} \footnote{ftp://legacy.gsfc.nasa.gov/caldb/data/asca/gis/cpf/95mar06/gis3v4\_0.rmf}.
The ancillary files were generated using the command, {\it arf}. The spectra were binned to achieve 25 counts in each bin.

X-ray spectral analysis was performed using the {\small XSPEC} fitting package, Version 12.8.1 \citep{1996ASPC..101...17A}. In this section,
we report the spectral parameters obtained by fitting XMM2.
 The total count rates in the source
and background regions for XMM2 were 0.02, 0.005, 0.004, 0.005, $7.0\times10^{-4}$ and $6.0\times10^{-4}$ c/s for the PN, MOS1 and MOS2 respectively.
 To account for the photoelectric absorption by the interstellar gas, two components were used. The first one was fixed to the Galactic value
 of $6\times10^{20}$ cm$^{-2}$ \citep{1990ARA&A..28..215D}. The other one was left free to account for the absorption within the SMC. For the latter component,
 the metal abundances were fixed at 0.2 solar, as is typical in the SMC \citep{1992ApJ...384..508R}.
  The atomic cross sections were adopted from \cite{1996ApJ...465..487V}. 
  
\begin{figure}
\centering
\includegraphics[scale=0.35]{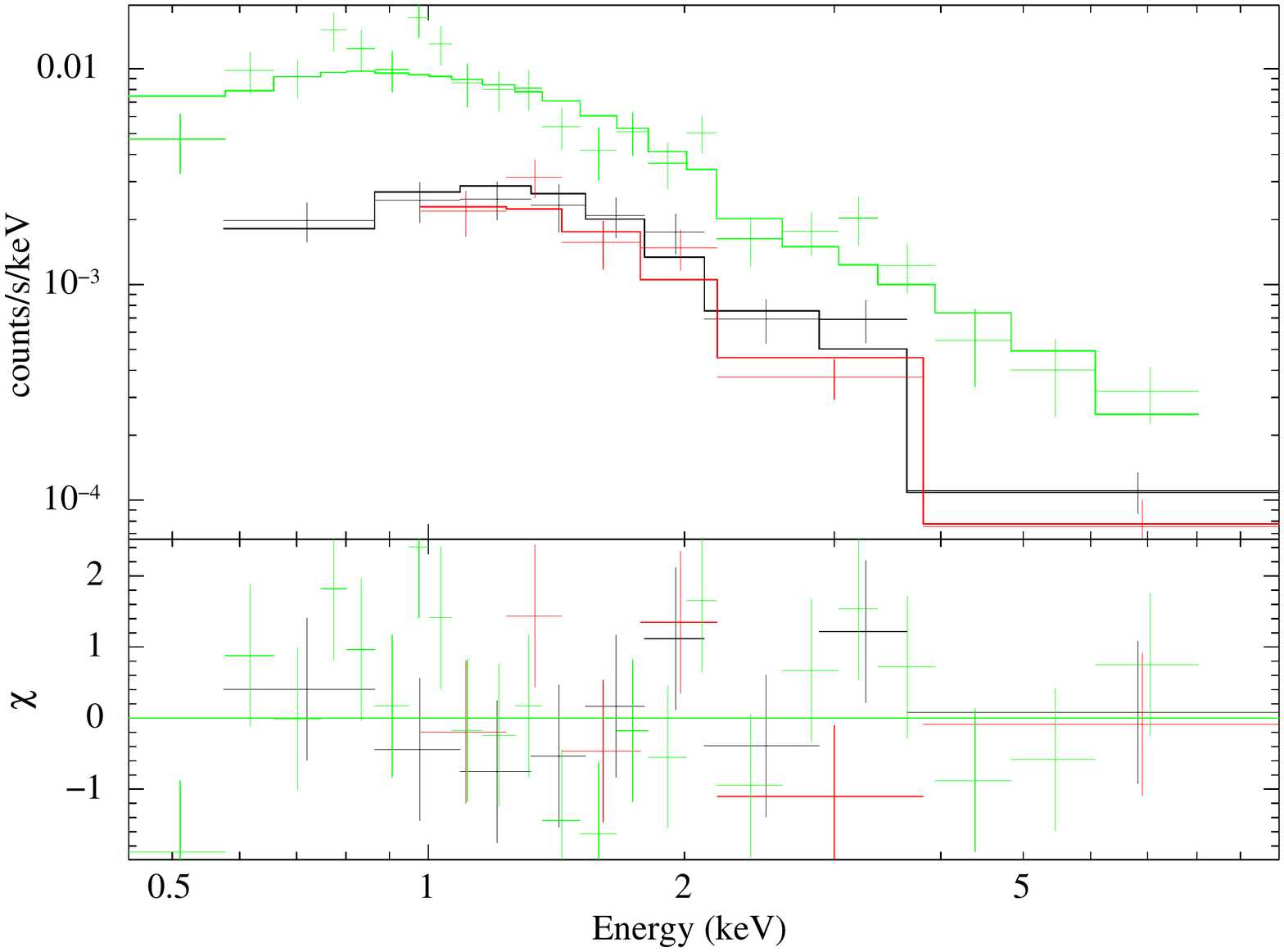}
\caption{The upper panel shows the simultaneous spectral fit of AX J0043$-$737 during XMM2 using spectra from all the 
\xmm/EPIC cameras (PN in `green', MOS1 in `red' and MOS2 in `black') along with the best-fit model. The lower panel displays the residuals after the fit.}
\label{spec}
\end{figure}

  An absorbed power law spectral model
provided an acceptable fit with reduced $\chi^{2}$ of 1.1 for 31 d.o.f (Fig. \ref{spec}). The spectral parameters are tabulated in Table~\ref{table2}.
The X-ray spectra of AGNs can typically be described by a power-law with photon index of $\Gamma\sim1.75$ \citep[e.g.][]{2006A&A...451..457T}.
The obtained $\Gamma$ of $\sim$ 1.7 is compatible with AX J0043$-$737 being an AGN behind the SMC. From the estimated luminosity-distance,
the determined absorption corrected X-ray luminosity (0.5--8.0 keV) is $1.9\times10^{44}$ erg s$^{-1}$.
  \begin{table}
\caption{Spectral parameters of AX J0043$-$737 and XMMU J004207.7$-$734503 from XMM2. Errors are quoted at 90\% confidence.}
\begin{tabular}{c c }
\hline \hline
Parameters & Value \\
\hline
 AX J0043$-$737 \\
 \hline
SMC $\ensuremath{N_{\mathrm{H}}}$ ($10^{22}$ atoms cm$^{-2}$) & 0.16 $\pm$ 0.07 \\
$\Gamma$ & 1.72 $\pm$ 0.19 \\
Flux $^a$ (0.5--8.0 keV) & $ 6.6 \pm 0.7 $\\
Flux (unabsorbed) $^a$ (0.5--8.0 keV) & $ 8.1 \pm 1.0 $\\
 \hline
 XMMU J004207.7$-$734503 \\
 \hline
 SMC $\ensuremath{N_{\mathrm{H}}}$ ($10^{22}$ atoms cm$^{-2}$) & 0.75 $\pm$ 0.34 \\
$\Gamma$ & 1.06 $\pm$ 0.16 \\
Flux $^a$ (0.5--8.0 keV) & $25.7 \pm 2.4$ \\
Flux (unabsorbed) $^a$ (0.5--8.0 keV) & $28.3 \pm 2.5 $\\
 \hline
\end{tabular}\\
$^{a}$ Flux in units of $10^{-14}$ \ergcm \\
\label{table2}
\end{table}
\subsection{Long term variability}
\label{sec:var}
The X-ray flux of AGNs varies on timescales ranging from hours to up to several years 
\citep[see e.g.][]{2012A&A...542A..83P,2011A&A...536A..84V}.
The three archival observations of the source (1 \xmm\, and 2 \asca) provide the opportunity to probe the long-term
 variability (if any) of the source spanning over 17 years. To test this we 
 performed spectral analysis as described below.

\begin{figure}
\centering
\includegraphics[scale=0.35]{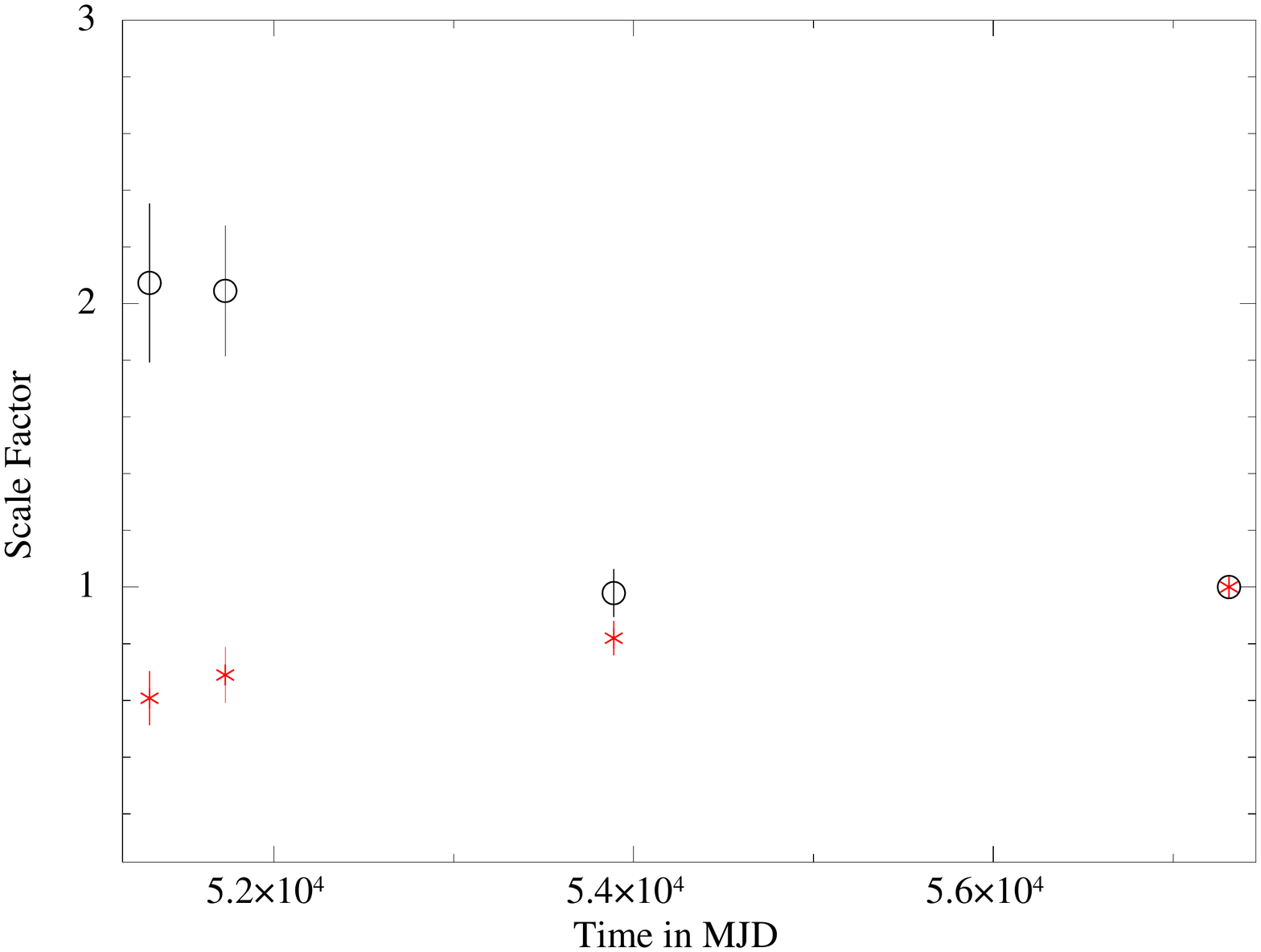}
\caption{Comparison of the scale factors (relative flux with respect to XMM2) for AX J0043$-$737 (in black circle) and XMMU J004207.7-734503 (in red cross)
with time in MJD.}
\label{fig2}
\end{figure}

 We fitted AX J0043$-$737 and XMMU J004207.7$-$734503 together because they are blended in the ASCA data. We used a two-component power-law model 
 (one corresponding to each source) and all the observations were fitted simultaneously. Each source was described by the same 
 spectral parameters in all observations. Possible variations were parameterised by scale factors with respect to XMM2 (relative flux with respect to the best observation, XMM2). 
 The other observations are not deep enough to test variability
 in the spectral shape and even if it existed, it would nearly always be associated with flux variability.
 For the \asca observations the fraction of spillover from one source into the other was taken into account. The difference in $\chi^{2}$
 with the scale factors fixed and left free is 40 for 3 d.o.f. This corresponds to a chance probability $< 0.00001$ and is highly significant.
 Fig.~\ref{fig2} shows a comparison of the variation of the scale factor over the observations for the two sources. AX J0043$-$737 shows
 a noticeable difference in the values especially between the \asca and \xmm\, observations. The corresponding values for 
 XMMU J004207.7-734503
 however do not show such a trend indicating that the variation is not due the differences in calibration between the two instruments.
 AX J0043$-$737 therefore exhibits signature of long-term variability typical for AGNs.
 
\section{Summary \& Conclusions}
\label{sec:conclusion}
With a dedicated \xmm\, observation of AX J0043$-$737 we classify it as an AGN behind the SMC.
The source coincides with MQS J004241.66-734041.3 which lies at a redshift of 0.95. The fit of a power-law model to the X-ray spectrum gives a spectral 
index typical for AGNs and has an absorption corrected X-ray luminosity of $1.9\times10^{44}$ erg $s^{-1}$ in the energy range of 0.5--8 keV.
By comparing the archival observations spanning over 17 years we find signatures of long-term variability. All of these are consistent
with the source being an AGN behind the SMC. More importantly, this answers the decade-long question on the nature of the source and 
rectifies its misidentification
as an 87 ms pulsar in the SMC.

\begin{acknowledgements}
The {\it XMM-Newton} project is supported by the Bundesministerium f{\"u}r Wirtschaft und Technologie/Deutsches Zentrum f{\"u}r Luft-und Raumfahrt (BMWi/DLR, FKZ 50 OX 0001) and the Max-Planck Society.
PE acknowledges funding in the framework of the NWO Vidi award A.2320.0076. We thank the referee for useful suggestions which significantly helped to improve the paper.
\end{acknowledgements}

\bibliographystyle{aa}
\bibliography{j0043}

\Online

\end{document}